\documentclass[useAMS,usenatbib]{mn2e}
\usepackage{epsfig}
\usepackage{times}

\newcommand{\MSun}{\mbox{${\rm M}_\odot$}}
\newcommand{\Msun}{\mbox{${\rm M}_\odot$}}

\newcommand{\RSun}{\mbox{${\rm R}_\odot$}}
\newcommand{\Rsun}{\mbox{${\rm R}_\odot$}}

\newcommand{\MJupiter}{\mbox{${\rm M}_{\rm Jup}$}}

\def\apgt{\ {\raise-.5ex\hbox{$\buildrel>\over\sim$}}\ }
\def\aplt{\ {\raise-.5ex\hbox{$\buildrel<\over\sim$}}\ }
\def\lt{\ {\raise-.5ex\hbox{$\buildrel>$}}\ }
\def\gt{\ {\raise-.5ex\hbox{$\buildrel<$}}\ }
\def\eqgt{\ {\raise-.5ex\hbox{$\buildrel>\over-$}}\ }
\def\eqlt{\ {\raise-.5ex\hbox{$\buildrel<\over-$}}\ }

\newfont{\Giga}{cmssbx10 scaled 5200}
\newfont{\giga}{cmssbx10 scaled 4300}
\newfont{\Mega}{cmssbx10 scaled 3200}
\newfont{\mega}{cmssbx10 scaled 2500}
\newfont{\Kilo}{cmssbx10 scaled 2000}
\newfont{\kilo}{cmssbx10 scaled 1600}
\newfont{\Deca}{cmssbx10 scaled 1450}
\newfont{\deca}{cmssbx10 scaled 1200}
\newfont{\Dezi}{cmssbx10 scaled 1100}
\newfont{\dezi}{cmssbx10 scaled 1050}
\newfont{\iGiga}{cmssi10 scaled 6200}
\newfont{\igiga}{cmssi10 scaled 4300}
\newfont{\iMega}{cmssi10 scaled 3200}
\newfont{\imega}{cmssi10 scaled 2500}
\newfont{\iKilo}{cmssi10 scaled 2000}
\newfont{\ikilo}{cmssi10 scaled 1500}
\newfont{\mathGiga}{cmsy10 scaled 6200}
\newfont{\mathgiga}{cmsy10 scaled 4300}
\newfont{\mathMega}{cmsy10 scaled 3200}
\newfont{\mathmega}{cmsy10 scaled 2500}
\newfont{\mathKilo}{cmsy10 scaled 2000}
\newfont{\mathkilo}{cmsy10 scaled 1500}
\newfont{\mathDeca}{cmsy10 scaled 1450}
\newfont{\mathdeca}{cmsy10 scaled 1200}

\def\apgt{\ {\raise-.5ex\hbox{$\buildrel>\over\sim$}}\ }
\def\aplt{\ {\raise-.5ex\hbox{$\buildrel<\over\sim$}}\ }

\title[Planet-Mediated Precision-Reconstruction of HU Aqr]{Planet-Mediated Precision-Reconstruction of the Evolution of the Cataclysmic Variable HU~Aquarius}
\author[S. Portegies Zwart]{
 S. Portegies Zwart \\
 Leiden Observatory, Leiden University, PO Box 9513,
 2300 RA, Leiden, The Netherlands 
}

\begin{document}

\date{}


\maketitle

\begin{abstract}

Cataclysmic variables (CVs) are binaries in which a compact white
dwarf accretes material from a low-mass companion star.  The discovery
of two planets in orbit around the CV HU Aquarii opens unusual
opportunities for understanding the formation and evolution of this
system.  In particular the orbital parameters of the planets
constrains the past and enables us to reconstruct the evolution of the
system through the common-envelope phase.  During this dramatic event
the entire hydrogen envelope of the primary star is ejected, passing
the two planets on the way.  The observed eccentricities and orbital
separations of the planets in HU~Aqr enable us to limit the
common-envelope parameter $\alpha \lambda = 0.45\pm 0.17$ or $\gamma =
1.77\pm0.02$ and measure the rate at which the common envelope is
ejected, which turns out to be copious.  The mass in the common
envelope is ejected from the binary system at a rate of ${\dot m} =
1.9\pm 0.3\,\MSun/yr$.  The reconstruction of the initial conditions
for HU~Aqr indicates that the primary star had a mass of $M_{\rm ZAMS}
= 1.6\pm0.2$\,\MSun\, and a $m_{\rm ZAMS} = 0.47\pm 0.04$\,\MSun\,
companion in a $a=25$--160\,\RSun\, (best value $a=97$\,\RSun)
binary. The two planets were born with an orbital separation of
$a_a=541\pm44$\,\RSun\, and $a_b=750\pm72$\,\Rsun\, respectively.
After the common envelope, the primary star turns into a
$0.52\pm0.01$\,\MSun\, helium white dwarf, which subsequently accreted
$\sim 0.30$\,\MSun\, from its Roche-lobe filling companion star,
grinding it down to its current observed mass of $0.18\,\MSun$.

\end{abstract}


\section{Introduction}

The cataclysmic variable HU Aqr currently consists of a 0.80\,\Msun\,
white dwarf that accretes from a 0.18\,\Msun\, main-sequence companion
star \citep{2011A&A...531A..34S}. The transfer of mass in the tight
$a=0.82$\,\RSun\, orbit is mediate by the emission of gravitational
waves and the strong magnetic field of the accreting star
\citep{1981A&A...100L...7V}.  Since its discovery
\citep{1993A&A...271L..25S}, irregularities of the observed-calculated
variations have led to a range of explanations, including the presence
of circum-binary planets
\citep{2009A&A...496..833S,2012arXiv1201.5730H,2012MNRAS.425..930G}.
Detailed timing analysis has eventually led to the conclusion that the
CV is orbited by two planets \citep{2012MNRAS.420.3609H}, a
5.7\,\MJupiter\, planet in a $\sim 1205$\,\RSun\, orbit with an
eccentricity of $e=0.20$ and a somewhat more massive (7.6\,\MJupiter)
planet in a wider $1785$\,\RSun\, and eccentric $e= 0.38$ orbit
\citep{2012arXiv1201.5730H}. Although, the two-planet configuration
turned out to be dynamically unstable on a 1000---10,000~year time
scale \citep[see also \S\,\ref{Sect:Stability}]{2012arXiv1201.5730H},
a small fraction of the numerical simulations exhibit long term
dynamical stability \cite[for model B2 in][see Tab.\,\ref{Tab:HUAqr}
  for the parameters]{2012MNRAS.419.3258W}.

It is peculiar to find a planet orbiting a binary, in particular
around a CV. While planets may be a natural consequence of the
formation of binaries \citep{PPZ2012}, planetary systems orbiting CVs
could also be quite common. In particular because of recently timing
residual in NN~Serpentis
\citep{2010A&A...521L..60B,2012MNRAS.425..749H}, DP Leonis
\citep{2011A&A...526A..53B} and QS Virgo \citep{2011IAUS..276..495A}
were also interpreted is being caused by circum-CV planets.

Although the verdict on the planets around HU Aqr (and the other CVs)
remains debated \citep[Tom Marsh private communication,
  and][]{2010MNRAS.401L..34Q}, we here demonstrate how a planet in
orbit around a CV, and in particular two planets, can constrain the CV
evolution and be used to reconstruct the history of the inner binary.
We will use the planets to perform a precision reconstruction of the
binary history, and for the remaining paper we assume the planets to
be real.

Because of their catastrophic evolutionary history, CVs seem to be the
last place to find planets.  The original binary lost probably more
than half its mass in the common-envelope phase, which causes the
reduction of the binary separation by more than an order of magnitude.
It is hard to imagine how a planet (let alone two) can survive such
turbulent past, but it could be a rather natural consequence of the
evolution of CVs, and its survival offers unique diagnostics to
constrain the origin and the evolution of the system.

\section{The evolution of a CV with planets}

After the birth of the binary, the primary star evolved until it
overflowed it Roche lobe, which initiated a common-envelope phase.
The hydrogen envelope of the primary was ejected quite suddenly in
this episode \citep{1984ApJ...277..355W}, and the white dwarf still
bears the imprint of its progenitor: the mass and composition of the
white dwarf limits the mass and evolutionary phase of its progenitor
star at the moment of Roche-lobe overflow (RLOF).  For an isolated
binary the degeneracy between the donor mass at the moment of RLOF
($M_{\rm RLOF}$), its radius $R_{\rm RLOF}$ and the mass of its core
$M_{\rm core}$ cannot be broken. 

The presence of the inner planet in orbit around HU Aqr
\citep{2012MNRAS.420.3609H,2012arXiv1201.5730H,2012MNRAS.425..930G}
allows us to break this degeneracy and derive the rate of mass loss in
the common-envelope phase. The outer planet allows us to validate this
calculation and in addition to determine the conditions under which
the CV was born. The requirement that the initial binary must have
been dynamically stable further constrains the masses of the two stars
and their orbital separation.

\subsection{Pre common envelope evolution}

During the CV phase little mass is lost from the binary system $M_{\rm
  cv} \simeq $\,constant \citep[but see][]{2002ASPC..261..242S}, and
the current total binary mass ($M_{\rm cv} = 0.98$\,\Msun) was not
affected by the past (and current) CV evolution
\citep{2010MmSAI..81..849R}. The observed white dwarf mass then
provides an upper limit to the mass of the core of the primary star at
the moment of Roche-lobe contact, and therefore also provides a
minimum to the companion mass via $m_{\rm comp} \eqgt M_{\rm cv} -
M_{\rm core}$.

With the mass of the companion not being affected by the common
envelope phase, we constrain the orbital parameters at the moment of
RLOF by calculating stellar evolution tracks to measure the core mass
$M_{\rm core}$ and the corresponding radius $R(M_{\rm core})$ for
stars with zero-age main-sequence mass $M_{\rm ZAMS}$.  In
Fig.\,\ref{fig:aMcoreForMzams3MSun} we present the evolution of the
radius of a 3\,\Msun\, star as a function of $M_{\rm core}$, which is
a measure of time.

We adopted the Henyey stellar evolution code MESA
\citep{2011ApJS..192....3P} to calculate evolutionary track of stars
from $M_{\rm ZAMS} = 1$ to 8\,\MSun\, using AMUSE\footnote{The
  Astrophysics Multipurpose Software Environment, or AMUSE, is a
  component library with a homogeneous interface structure, and can be
  downloaded for free at {\tt amusecode.org}. All the source codes and
  scripts for reproducing the calculations described in this paper are
  available via the AMUSE website.}
\citep{2009NewA...14..369P,2012arXiv1204.5522P} to run MESA and
determine the mass of the stellar core. The latter is measured by
searching for the mass-shell in the stellar evolution code for which
the relative Hydrogen fraction $<1.0^{-9}$
\citep{2001A&A...369..170T}.

At the moment of RLOF the core mass is $M_{\rm core}$ and the stellar
radius $R_{\rm RLOF} \equiv R(M_{\rm core})$.  Via the relation for
the Roche radius~\citep{1983ApJ...268..368E}, we can now calculate the
orbital separation at the moment of RLOF $a_{\rm RLOF}$ as a function
of $M_{\rm RLOF}$. This separation is slightly larger than the initial
(zero-age) binary separation $a_{\rm ZAMS}$ due to the mass lost by
the primary star since its birth $M_{\rm RLOF}-M_{\rm ZAMS}$.  The
long (main-sequence) time scale in which this mass is lost guarantees
an adiabatic response to the orbital separation, i.e.\, $a M_{\rm tot}
=$~constant.

For each $M_{\rm ZAMS}$ we now have a range of possible solutions for
$a_{\rm RLOF}$ as a function of $M_{\rm core}$ and $m_{\rm comp} =
M_{\rm cv} - M_{\rm core}$.  This reflects the assumption that the
total mass ($m_{\rm comp} + M_{\rm core} = 0.98\,\MSun$) in the
observed binary with mass $M_{\rm cv}$ is conserved throughout the
evolution of the CV.  In Fig.\,\ref{fig:aMcoreForMzams3MSun} we
present the corresponding stellar radius $R_{\rm RLOF}$ and $a_{\rm
  RLOF}$ as a function of $M_{\rm core}$ for $M_{\rm ZAMS}=3\,\MSun$.
This curve for $a_{\rm RLOF}$ is interrupted when RLOF would already
have been initiated earlier for that particular orbital separation.
We calculate this curve by first measuring the size of the donor for
core mass $M_{\rm core}$, and assuming that the primary fills its
Roche-lobe we calculate the orbital separation at which this happens.

\subsection{The common envelope evolution}\label{Sect:CE}

During the common envelope phase the primary's mantle is blown away
beyond the orbit of the planets. The latter responds to this by
migrating from the orbits in which they were born (semi-major axis
$a_a$ and eccentricity $e_a$, the subscript '$a$' indicates the inner
planet, we adopt a '$b$' to indicate the outer planet) to the
currently observed orbits.  Using first order analysis we recognize
two regimes of mass loss: fast and slow. In the latter case the orbit
expands adiabatically without affecting the eccentricity: The minimum
possible expansion of the planet's orbit is achieved when the common
envelope is lost adiabatically. Fast mass loss leads to an increase in
the eccentricity as well and may even cause the planet to escape
\citep{1983ApJ...267..322H,2012MNRAS.424.2914P}.

\begin{figure}
 \centering
 \psfig{file=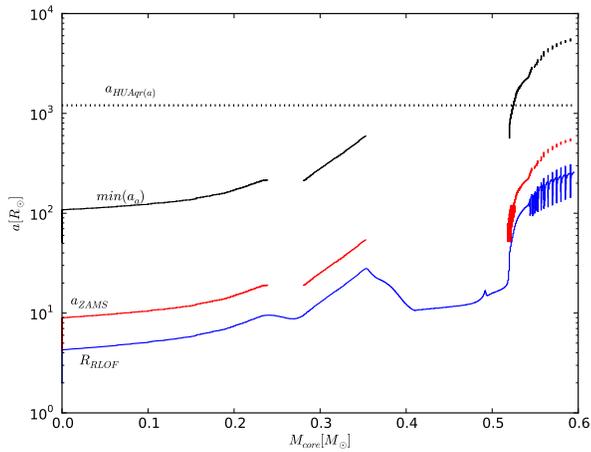,width=0.5\textwidth}
 \caption{The orbital separation as a function of core mass for a
   primary star of $M_{\rm ZAMS} = 3$\,\MSun. The thin blue curve
   (bottom) gives the stellar radius $R_{\rm RLOF}$ as a function of
   its core mass $M_{\rm core}$, and the curve directly above that
   (red) gives the initial orbital separation $a_{\rm ZAMS}$ for which
   Roche-lobe overflow of that binary occurs. Here we adopted a
   companion mass of $m_{\rm comp} = 0.98\MSun-M_{\rm core}$ as in HU
   Aqr.  The interruptions in the curves indicate the core masses for
   which Roche-lobe overflow cannot occur, because it would already
   have occurred in an earlier stage of the evolution, i.e., at a
   smaller core mass. The solid black curve (top) gives the minimal
   orbital separation of a planet born at $a_a=3a_{\rm ZAMS}$ for with
   the orbit was adiabatically expanded due to the mass loss in the
   common envelope $M_{\rm ZAMS}-M_{\rm core}$.  The horizontal dotted
   curve gives the separation at which the inner planet around HU Aqr
   was observed.  For viable solutions the solid black curve should
   remain below the horizontal dotted curve.  The thick parts of the
   red curve indicate where the zero-age binary complies to the most
   favorable conditions for engaging RLOF, surviving the
   common-envelope and produce a planet that can migrate to at least
   the observed separation for the inner planet in HU Aqr.  }
 \label{fig:aMcoreForMzams3MSun}
\end{figure}

A planet born at the shortest possible orbital separation to be
dynamically stable will have $a_a \sim 3a_{\rm ZAMS}$
\citep{2001MNRAS.321..398M}, which is slightly smaller than the
distance at which circum binary planets tend to form \citep{PPZ2012}.
In Fig.\,\ref{fig:aMcoreForMzams3MSun} we present a minimum to the
semi-major axis for a planet that was born at $a_a = 3a_{\rm ZAMS}$
and migrated by the adiabatic loss of the hydrogen envelope from the
primary star in the common-envelope phase.  The planet can have
migrated to a wider orbit, but not to an orbit smaller than the solid
black curve (indicated with $min(a_a)$) in
Fig.\,\ref{fig:aMcoreForMzams3MSun}.  For the 3\,\MSun\, star,
presented in Fig.\,\ref{fig:aMcoreForMzams3MSun}, RLOF can
successfully result in the migration of the planet to the observed
separation in HU Aqr for $M_{\rm core} \aplt 0.521$\,\RSun, which
occurs for $a_{\rm ZAMS} \aplt 111$\,\RSun.  A core mass $M_{\rm core}
> 0.521$\,\RSun\, would, for a 3\,\Msun\, primary star, result in an
orbital separation that exceeds that of the inner planet in HU Aqr; in
this case the core mass of the primary star must have been smaller
than 0.521\,\MSun.

Another constraint on the initial binary orbit is provided by the
requirement that the mass transfer in the post common-envelope binary
should be stable when the companion starts to overfill its Roche lobe.
To guarantee stable mass transfer we require that $m_{\rm comp} \apgt
M_{\rm core}$.  The thick part of the red curve in
Fig.\,\ref{fig:aMcoreForMzams3MSun} indicates the valid range for the
initial orbital separation and core-mass for which the observed planet
can be explained; the thin parts indicate where these criteria fail.

\begin{figure}
 \centering 
 \psfig{file=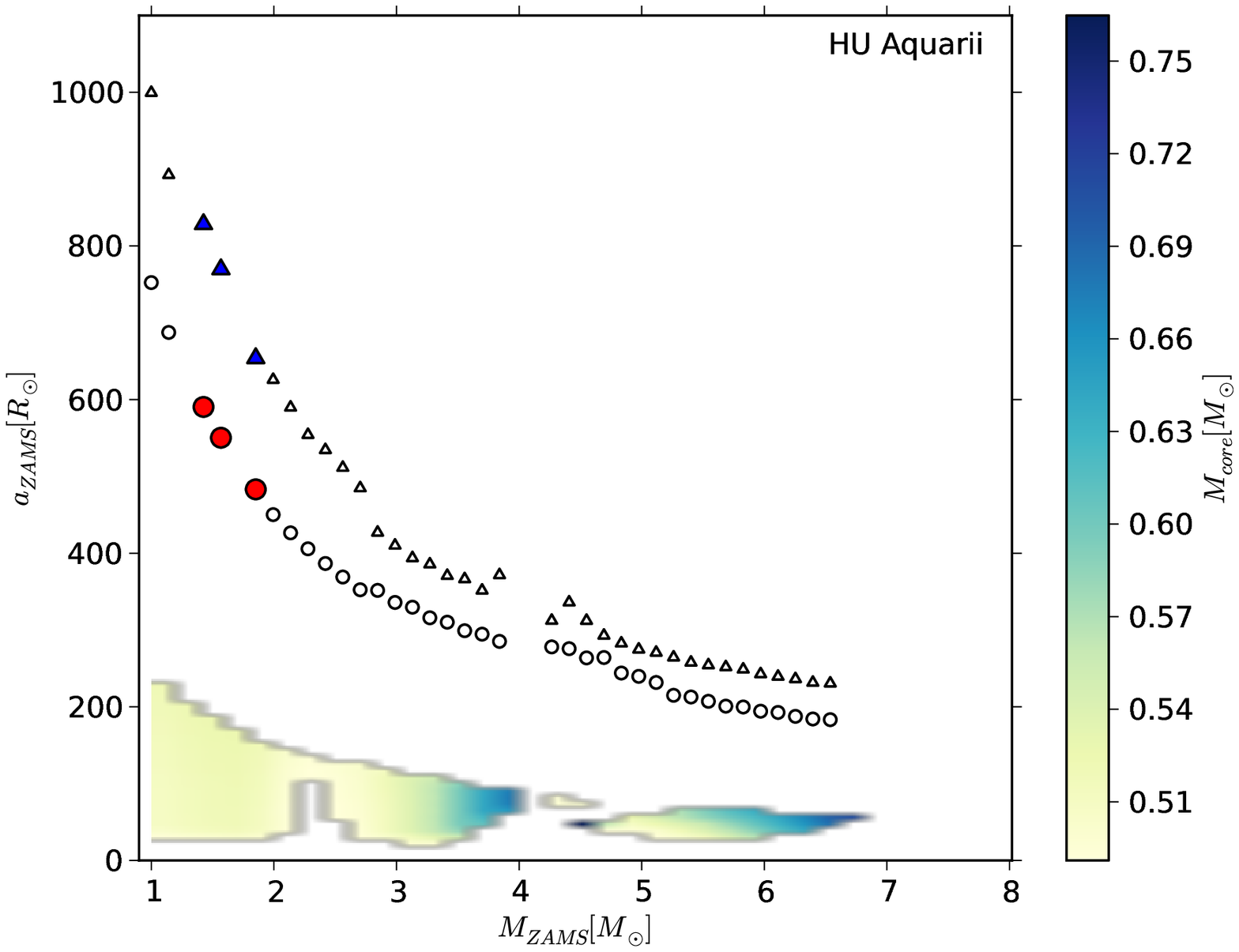,width=0.5\textwidth}
 \psfig{file=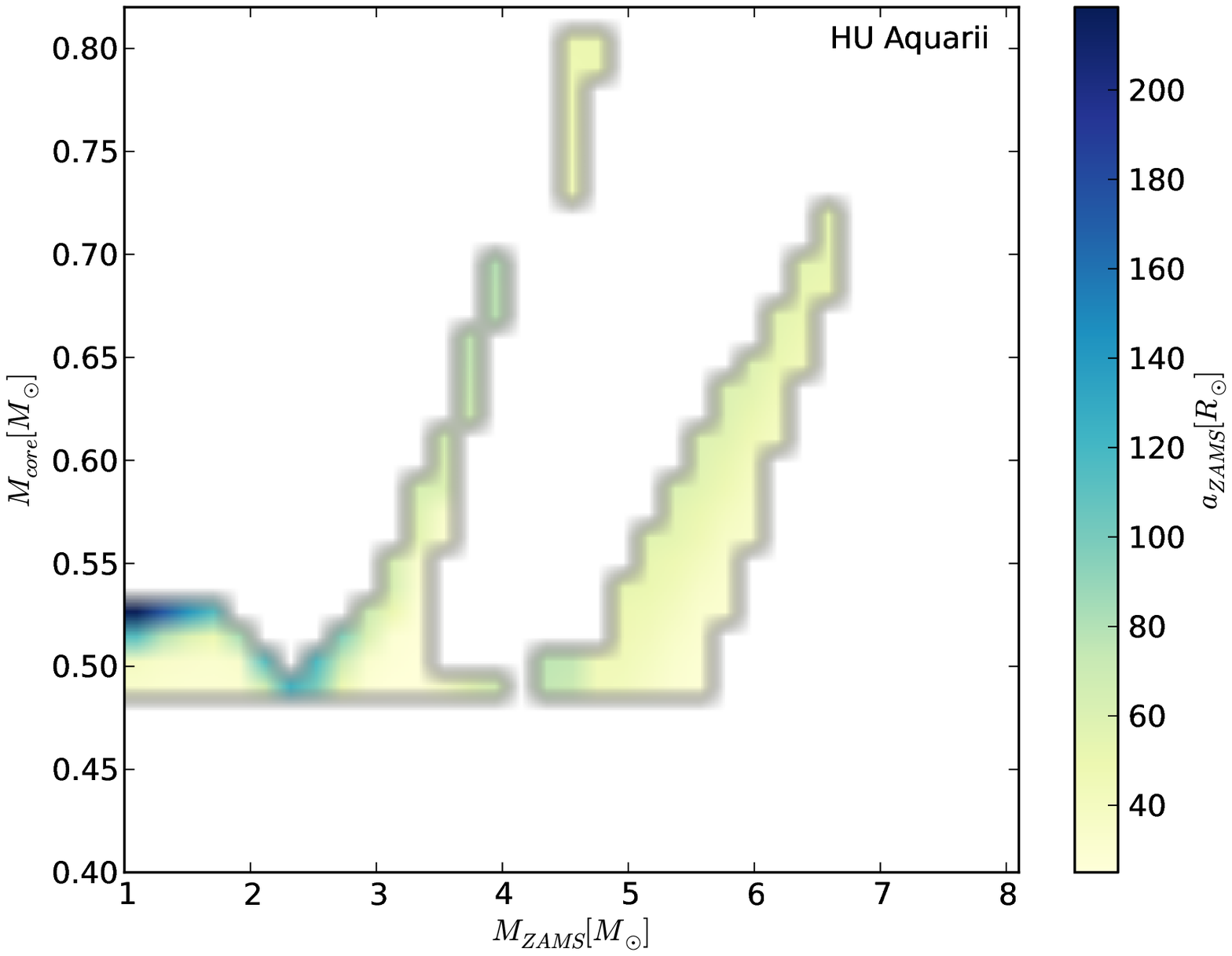,width=0.5\textwidth}
 \caption{Distribution of initial conditions $a_{\rm ZAMS}$ and
   $M_{\rm ZAMS}$, and the resulting core mass $M_{\rm core}$ (bottom
   panel) that successfully reproduce the CV HU Aqr.  In the top panel
   we also present the results of our analysis for the two
   planets. The circles give the initial semi-major axis of the inner
   planet. The triangles give the initial semi-major axis for the
   outer planet. The symbols are colored (red for the inner planet and
   blue for the outer) if at least 10 out of 20 calculations for the
   orbital integration over 1\,Myr turn out to be dynamically stable
   (see\,\ref{Sect:Stability}).  With these initial conditions both
   planets migrate to within 1\% of their observed orbital period with
   an eccentricity of $e_a=0.20\pm0.01$ and $e_b=0.33\pm0.03$ with
   ${\dot m}_a = 0.15\pm0.01$\,\MSun/yr.}
 \label{fig:aM0_distribution_HU}
\end{figure}

We repeat the calculation presented in
Fig.\,\ref{fig:aMcoreForMzams3MSun} for a range of masses from $M_{\rm
  ZAMS} = 1$\,\Msun\, to 8\,\MSun\, with steps of 0.02\,\MSun, the
results are presented as the shaded region in
Fig.\,\ref{fig:aM0_distribution_HU}.

\section{Effect of the common envelope phase on the planetary system}

The response of the orbit of the planet to the mass loss depends on
the total amount of mass lost in the common envelope and the rate at
which it is lost.  Numerical common-envelope studies indicate that for
an in-spiraling binary ${\dot m} \simeq 2.0\,\MSun/yr$
\citep{2012ApJ...746...74R}.  At this rate the entire envelope $M_{\rm
  RLOF} - M_{\rm core} \sim 0.46$--5.8\,\MSun\, is expelled well
within one orbital period of the inner planet, which leads to an
impulsive response and the possible loss (for $M_{\rm RLOF} - M_{\rm
  core} \apgt 1.92$\,\MSun) of the planet.  The fact that the HU Aqr
is orbited by a planet indicates that at the distance of the planet
${\dot m}_a \ll (M_{\rm RLOF} - M_{\rm core})/P_{\rm planet} \sim
1\,\MSun/yr$. The eccentricity of the inner planet in HU Aqr (see
Tab.\,\ref{Tab:HUAqr}) can be used to further constrain the rate at
which the common-envelope was lost from the planetary orbit. The
higher eccentricity of the outer planet indicates a more impulsive
response, which is a natural consequence of its wider orbits with the
same ${\dot m}$. This regime between adiabatic and impulsive mass loss
is hard to study analytically \citep{2008ARep...52..806L}.

\begin{table*}
\centering
\caption{ Reconstructed and observationally constrained parameters for
  HU Aquarii.  The parameters at zero-age (column 2 and 3) and
  directly after the common-envelope phase (columns 4 and 5) are
  derived by means of reconstructing the CV evolution.  The best
  comparison is achieved for a mass loss rate in the common-envelope
  from the binary system of ${\dot m} = 1.9\pm 0.3\,\MSun/yr$ (or
  ${\dot m}_a = 0.15\pm0.01$\,\MSun/yr from the orbit of the first
  planet).  Parameters that we were unable to constrain are printed in
  slanted font, observed parameters are from
  \citep{2011A&A...531A..34S,2012MNRAS.420.3609H,2012MNRAS.425..930G}
  and are printed in boldface.
  \label{Tab:HUAqr}
}
\begin{tabular}{llllllll}
parameter     & \multicolumn{2}{c}{Zero age}      & \multicolumn{2}{c}{After CE}     & \multicolumn{2}{c}{Today} \\  
\hline
\hline
$M/\MSun$     & \multicolumn{2}{c}{$1.6\pm0.2$} & \multicolumn{2}{c}{$0.52\pm0.01$}& \multicolumn{2}{c}{${\bf 0.80\pm0.04}$} \\
$m/\MSun$     & \multicolumn{2}{c}{$0.47\pm 0.04$}& \multicolumn{2}{c}{$0.47\pm0.04$}& \multicolumn{2}{c}{${\bf 0.18\pm0.06}$} \\
$a/\RSun$     & \multicolumn{2}{c}{25--160}       & \multicolumn{2}{c}{$0.867$--2.0}  & \multicolumn{2}{c}{\bf 0.8} \\
$e$           & \multicolumn{2}{c}{\sl 0.0}      & \multicolumn{2}{c}{\sl 0.0}     & \multicolumn{2}{c}{\sl 0.0} \\
              & planet a   & planet b             & planet a    & planet b           & planet a    & planet b  \\         
$a/\RSun$     & $541\pm44$  & $750\pm72$          & $1204\pm12$ & $1785\pm18$        & {\bf 1204}        &  {\bf 1785}    \\
$m/\MJupiter$ & {\sl 5.7}   & {\sl 7}             & {\sl 5.7}   & {\sl 7}            & {\bf 5.7}       &   {\bf  7}    \\
$e$           & {\sl 0.0}   & {\sl 0.0}           &$0.20\pm0.00$& $0.33\pm0.03$      & {\bf 0.20}        &  {\bf 0.38}     \\
\hline
\hline
\end{tabular}
\end{table*}

\subsection{The response of the inner planet}\label{Sect:InnerPlanet}

We calculate the effect of the mass loss on the orbital parameters by
numerically integrating the planet orbit. The calculations are started
by selecting initial conditions for the zero-age binary HU Aqr
--$M_{\rm ZAMS}$, $a_{\rm ZAMS}$ and consequently $M_{\rm core}$--
from the available parameter space (shaded area) in
Fig.\,\ref{fig:aM0_distribution_HU}, and integrate the equations of
motion of the inner planet with time.  Planets ware assumed to be born
in a circular orbit ($e_a=0$) in the binary plane with semi-major axis
$a_a$.

The equations of motions are integrated using the high-order
symplectic integrator Huayno \citep{2012NewA...17..711P} via the
AMUSE framework. During the integration we adopt a constant mass-loss
rate ${\dot m}$ applied at every 1/100th of an orbit, and we continued
the calculation until the entire envelope is lost (see
\S\,\ref{Sect:CE} and Fig.\,\ref{fig:aM0_distribution_HU}), at which
time we measure the final semi-major axis and eccentricity of the
planetary orbit.  During the integration we allow the energy error to
increase up to at most $\Delta E/E = 10^{-13}$.

By repeating this calculation while varying $a_a$ and ${\dot m}_a$ we
iterate (by bisection) until the result is within 1\% of the observed
$a_{\rm HUAqr}(a)$ and $e_{\rm HUAqr}(a)$ of the inner planet observed
in HU Aqr.  The converged results of these simulations are presented
in Fig.\,\ref{fig:aM0_distribution_HU} (circles), and these represent
the range of consistent values for the inner planet's orbital
separation $a_a = 183$--752\,\RSun\, as a function of $M_{\rm
  ZAMS}=1$--8\,\MSun\, and consistently reproduce the observed inner
planet when adopting ${\dot m}_a = 0.124$--0.267\,\MSun/yr.  The highest
value for ${\dot m}$ is reached for $M_{\rm ZAMS} = 2.85$\,\MSun\, at
an initial orbital separation of $a_a = 427$\,\RSun.

The orbital solution for the inner planet is insensitive to the
semi-major axis of the zero-age binary $a_{\rm ZAMS}$ (for a fixed
$M_{\rm ZAMS}$), and each of these solutions were tested for dynamical
stability, which turned out to be the case irrespective of the initial
binary semi-major axis (as discussed in \S\,\ref{Sect:Stability}).

\subsection{The response of the outer planet}

We now adopt the in \S\,\ref{Sect:InnerPlanet} measured value of
${\dot m}$ to integrate the orbit of the outer planet.  The effect of
the mass outflow on the planet is proportional to the square of the
density in the wind at the location of the planet
\citep{1992A&A...257..655K}. We correct for this effect by reducing the
mass loss rate in the common envelope that affects the outer planet by
a factor $(a_b/a_a)^{3/2}$.

We use the same integrator and assumptions about the initial orbits as
in \S\,\ref{Sect:InnerPlanet}, but we adopt the value of ${\dot m}$
from our reconstruction of the inner planet (see
\S\,\ref{Sect:InnerPlanet}).  To reconstruct the initial orbital
separation of the outer planet $a_{b}$, we vary this value (by
bisection) until the final semi-major axis is within 1\%\, of the
observed orbit (see Tab.\,\ref{Tab:HUAqr}). The results are presented
in Fig.\,\ref{fig:aM0_distribution_HU} (triangles).  The post
common-envelope eccentricity of the outer planet then turn out to be
$e_b = 0.38\pm0.07$.

\section{Stability of the initial system}\label{Sect:Stability}

After having reconstructed the initial conditions of the binary system
with its two planets we test its dynamical stability by integrating
the entire system numerically for 1\,Myr using the Huayno integrator
\citep{2012NewA...17..711P}.  To test the stability we check the
semi-major axis and eccentricity of both planets every 100\,years. If
any of these parameters change by a factor of two compared to the
initial values or if the orbits cross we declare the system unstable,
otherwise they are considered stable.  The calculations are repeated
with the 4th order Hermite predictor-corrector integrator ph4
\citep{2012ASPC..453..129M} within AMUSE to verify that the results
are robust, which turned out to be the case.  We then repeated this
calculation ten times with random inital orbital phases and again with
a 1\% Gaussian variation in the initial planetary semi-major axes.  In
Fig.\,\ref{fig:aM0_distribution_HU} we present the resulting stable
systems by coloring them red (circled) and blue (triangles), the
unstable systems are represented by open symbols.

From the wide range of possible systems that can produce HU Aqr only a
small range around $M_{\rm ZAMS} = 1.6\pm0.2$\,\MSun\, turns out to be
dynamically stable.  The eccentricity of the outer orbit of the stable
systems (which ware stable for initial conditions within 1\%) $e_b =
0.32\pm 0.02$, which is somewhat smaller than the observed value for
HU Aqr \citep[$e = 0.38\pm 0.16$]{2012arXiv1201.5730H}.  These values
are obtained with ${\dot m}_a = 0.15\pm 0.01\,\MSun/yr$.  The small
uncertainty in the derived value of ${\dot m}$ is a direct consequence
of its sensitivity to $e_a$ and the small error on $M_{\rm ZAMS}$ from
the requirement that the initial system is dynamically stable.

\section{Discussion and conclusions}

We have adopted the suggestive results from the timing analysis of HU
Aqr, that the CV is orbited by two planets, to reconstruct the
evolution of this complex system.  A word of caution is well placed in
that these observations are not confirmed, and currently under debate
(Tom Marsh private communication, and comments by the
referee). However, The predictive power that such an observation would
entail is interesting. The possibility to reconstruct the initial
conditions of a CV by measuring the orbital parameters of two circum
binary planets is a general result that can be applied to other
binaries. For CVs in particular it enables us to constrain the value
of fundamental parameters in the common-envelope evolution. This in
itself makes it interesting to perform this theoretical exercise,
irrespective of the uncertainty in the observations.  On the other
hand, the consistency between the observations and the theoretical
analysis give some trust to the correctness of these observations.

The presence of one planet in an eccentric orbit around a CV allow us
to calculate the rate at which the common-envelope was lost from the
inner binary.  A single planet provides insufficient information to
derive the initial mass of the primary star, but allows us to derive
the initial binary separation and planetary orbital separation to
within about factor of 5, and the initial rate of mass loss from the
common envelope to about a factor 2.  A second planet can be used to
further constrain these parameters to a few per cent accuracy and
allows us to make a precision reconstruction of the evolution of the
CV.

We have used the observed two planets in orbit around the CV HU Aqr to
reconstruct its evolution, to derived its initial conditions (primary
mass, secondary mass, orbital separation, and the orbital separations
of both planets) and to measure the rate of mass lost in the
common-envelope parameters ${\dot m}$. By comparing the binary
parameters at birth with those after the common-envelope phase we
subsequently calculate the two parameters $\alpha\lambda$ and
$\gamma$.

The measured rate of mass loss for HU Aqr of ${\dot m}_a = 0.15\pm
0.01\,\MSun/yr$ from the inner planetary orbit, which from the binary
system itself would entail a mass-loss rate of ${\dot m} = 1.9\pm
0.3\,\MSun/yr$, when we adopt the initial binary to have a semi-major
axis of $a_{\rm ZAMS} \simeq 97$\,\RSun, which is bracketed by our
derived range of $a_{\rm ZAMS} = 25$--160\,\RSun.  This is consistent
with a mass-loss rate of ${\dot m} \simeq 2\,\MSun/yr$ from numerical
common-envelope studies \citep{2012ApJ...746...74R}.

By adopting that the binary survives its common envelope at a
separation between $\sim 0.87$\,\RSun\, (at which separation the
secondary star will just fill it's Roche-lobe to the white dwarf) and
$\sim 2$\,\RSun\, (for gravitational wave radiation to drive the
binary into Roche-lobe overflow within 10\,Gyr), we derive the value
of $\alpha \lambda = 0.2$--2.0 (for $a_{\rm ZAMS} \simeq 97$\,\RSun\,
we arrive at $\alpha \lambda \simeq 0.45\pm 0.17$). This value is a
bit small compared to numerous earlier studies, which tend to suggest
$\alpha \lambda \simeq 4.0$. The alternative $\gamma$-formalism for
common-envelope ejection \citep{2000A&A...360.1011N} gives a value of
$\gamma = 1.63$--1.80 (for $a_{\rm ZAMS} \simeq 97$\,\RSun\, we arrive
at $\gamma \simeq 1.77 \pm 0.02$), which is consistent with the
determination of $\gamma$ in 30 other CVs \citep{2005MNRAS.356..753N}.

The inner planet in HU Aqr formed at $(3.2$--$20.6)a_{\rm ZAMS}$, with
a best value of $5.3\pm0.45 a_{\rm ZAMS}$, which is consistent with
the planets found to orbit other binaries, like Kepler 16
\citep{2011Sci...333.1602D} and for Kepler 34 and 35
\citep{2012Natur.481..475W}, although these systems have lower primary
mass and secondary mass stars.

It seems unlikely that more planets were formed inside the orbit of
the inner most planet, even though currently there is sufficient
parameter space for many more stable planets; in the zero-age binary
there has not been much room for forming additional planets further
in. It is however possible that additional planets formed further out
and those, we predict, will have even higher eccentricity than those
already found.

{\bf Acknowledgements} It is a pleasure to thank Edward P.J.\, van den
Heuvel, Tom Marsh, Inti Pelupessy, Nathan de Vries, Arjen van Elteren
and the anonymous referee for comments on the manuscript and
discussions.  This work was supported by the Netherlands Research
Council NWO (grants \#612.071.305 [LGM], \#639.073.803 [VICI] and
\#614.061.608 [AMUSE]) and by the Netherlands Research School for
Astronomy (NOVA).

\input journalnames.sty


\end{document}